\documentclass[12pt]{article}

\usepackage{amsfonts,amssymb,amsbsy,latexsym,amsmath}

\usepackage[cp866]{inputenc}
\usepackage[russian]{babel}

\begin{document}

\renewcommand{\thefootnote}{\fnsymbol{footnote}}

\renewcommand{\arctg}{\mathop{\rm arctg}\nolimits}
\renewcommand{\sh}{\mathop{\rm sh}\nolimits}
\renewcommand{\ch}{\mathop{\rm ch}\nolimits}
\renewcommand{\th}{\mathop{\rm th}\nolimits}
\renewcommand{\ctg}{\mathop{\rm ctg}\nolimits}
\renewcommand{\tg}{\mathop{\rm tg}\nolimits}
\newcommand{\grad}{\mathop{\rm grad}\nolimits}
\newcommand{\rot}{\mathop{\rm rot}\nolimits}
\newcommand{\arcsh}{\mathop{\rm arcsh}\nolimits}
\newcommand{\tot}{\mathop{\rm tot}\nolimits}

\def\s#1{\sqrt{#1}}

\newcommand{\be}{\begin{equation}}
\newcommand{\ee}{\end{equation}}
\newcommand{\ba}{\begin{eqnarray}}
\newcommand{\ea}{\end{eqnarray}}
\newcommand{\pa}{\partial}
\let\f\frac
\newcommand{\st}{\stackrel}
\newcommand{\tk}{\tilde\kappa}
\newcommand{\ep}{\epsilon}
\newcommand{\ds}{\displaystyle}

\large
\def\g{\gamma\,'}
\def\gR{\mathfrak{R}}
\def\gv{\mathfrak{v}}
\def\gc{\mathfrak{c}}
\def\gr{\mathfrak{r}}
\def\be{\begin{equation}}
\def\ee{\end{equation}}
\def\ba{\begin{eqnarray}}
\def\ea{\end{eqnarray}}
\def\p{\partial}
\def\f{\frac}
\def\s{\sqrt}
\def\gvmn{g^{\mu\nu}}
\def\gnmn{g_{\mu\nu}}
\def\gavmn{\gamma^{\mu\nu}}
\def\ganmn{\gamma_{\mu\nu}}
\def\vnmn{\varphi_{\mu\nu}}
\def\S{Schwarzschild}

\begin{center}
{\large\bf ON THE INTERNAL SOLUTION OF THE SCHWARZSCHILD TYPE IN THE
FIELD THEORY
OF GRAVITATION}

\vspace*{0.2cm}

{\large \bf  S.S. Gershtein, A.A. Logunov\footnote{e-mail:
Anatoly.Logunov@ihep.ru}, M.A. Mestvirishvili}

\vspace*{0.4cm}

{ \it Institute for High Energy Physics, Protvino, Russia }

\vspace*{0.4cm}

\end{center}

\centerline{\bf Abstract}

\vspace*{0.4cm}
{\small It is shown  that the internal solution of the
Schwarzschild type in the Relativistic Theory of Gravitation does not
lead to an {infinite pressure} inside a body as it holds in the
General Theory of Relativity. 

This happens due to the graviton rest mass, because  of the  stopping
of
the time slowing down. 
}
\vspace*{0.4cm}

K. Schwarzschild found in papers [1,2]  a spherically symmetric
static
solution (internal and external) of the general relativity (GR)
equations. The external solution is widely known and has the
following form:
\be
ds^2=c^2\Bigl(1-\f{W_g}{W}\Bigr)dt^2
-\Bigl(1-\f{W_g}{W}\Bigr)^{-1}dW^2
+W^2(d\theta^2+\sin^2\theta\,d\phi^2)\,,
\label{eq1}
\ee
where  $W_g=(2GM)/c^2$ is the \S~radius.

The internal \S~solution for a {\bf
ho\-mo\-ge\-ne\-ous
ball}
of the radius $a$ is described by the interval: 
\ba
\begin{cases}
ds^2=c^2\Bigl(\ds\f{\,3\,}{2}\sqrt{1-q a^2}-\f{\,1\,}{2}\sqrt{1-q
W^2}\Bigr)^2dt^2-\\[4mm]
\quad\;\;\,-(1-qW^2)^{-1}dW^2
 +W^2(d\theta^2+\sin^2\theta\,d\phi^2)\,,
\end{cases}
\label{eq2}
\ea
where~~$q=(1/3)\varkappa\rho =(2GM)/(c^2 a^3)$,~~ $\varkappa = (8\pi
G)/c^2$, \\
$\rho =(3M)/(4\pi a^3)$.

The general property of internal and external solutions is
ma\-ni\-fes\-ted 
in the fact that at a certain value of $W$ the metric coefficients in
front of
$dt^2$ in intervals (1) and (2) vanish. Vanishing of this metric
coefficient, which we will denote as $U$, means that the
gravitational
field acts in such a
way that it not only slows down the run of time, but is able even
{to stop this run}. For the external solution the vanishing of the
metric coefficient $U$ occurs at $W=W_g$. 

To exclude such a possibility (not forbidden by the theory) one has
to
assume that the radius of the body obeys to the inequality
\be
a>W_g\,.
\label{eq3}
\ee
For the internal solution this occurs at 
\be
W^2 = 9a^2 - 8(a^3/W_g)\,,
\label{eq4}
\ee
To exclude such a possibility of the vanishing of the metric
coefficient
$U$ inside the body one has to assume that 
\be
a>(9/8)W_g\,.
\label{eq5}
\ee
\textit{We have to emphasize that inequalities (\ref{eq3}) and
(\ref{eq5}) 
are not a consequence of GR.}

The internal \S~ solution is somewhat formal and is interesting first
of all
because it is an exact solution of the GR equations. In papers [3,4]
it was shown, taking as an example the external \S~ solution that
in
the Relativistic Theory of Gravitation (RTG), as a field
theory,
inequality (3) arises due to an effective repulsive force, which is
stipulated by the stop of the time slowing down, which, in its turn,
is caused by the graviton rest mass. Below we consider, 
in the framework of the RTG, the internal solution of the \S~ type. 
The internal Schwarzschild 
solution is the solution of the Hilbert--Einstein
equ\-a\-ti\-ons
\[
1-\ds\f{d}{dW}\Bigl[\f{W}{V}\Bigr]=\varkappa W^2\rho\,,
\]
\be
1-\f{1}{V}-\f{W}{UV}\f{dU}{dW}
=-\varkappa \f{W^2}{c^2}p\,.
\label{eq6}
\ee
According to (\ref{eq2}) the metric coefficients in front of $dt^2$
and $dW^2$ are, respectively, 
\be
U=\Bigl(\f{\,3\,}{2}\sqrt{1-q a^2}
-\f{\,1\,}{2}\sqrt{1-q W^2}\Bigr)^2,\quad
V=(1-q W^2)^{-1}.
\label{eq7}
\ee
We find thereof
\be
\f{\overset{\;\,\prime}{U}}{U}
=\f{q W}{\sqrt{1-q W^2}\Bigl(\ds\f{\,3\,}{2}\sqrt{1-q a^2}
-\ds\f{\,1\,}{2}\sqrt{1-q W^2}\Bigr)}\,,\quad
{\overset{\;\,\prime}{U}}=\f{dU}{dW}\,.
\label{eq8}
\ee
Substituting (\ref{eq7}) and (\ref{eq8}) into equation  (\ref{eq6}),
we obtain the pressure
\be
\f{p}{c^2}
=\f{\,\rho\,}{2}\ds\f{(\sqrt{1-q W^2}-\sqrt{1-q a^2})}{\sqrt{U}}\,.
\label{eq9}
\ee 

It is seen from this, in particular, that if equation (\ref{eq4})
would not be excluded then the pressure inside the body would be
infinite on
the sphere defined by this equation. The singularity that arises
due to the vanishing of the metric coefficient $U$ cannot be
eliminated
by the choice of the coordinate system, because the scalar curvature
$R$ also possesses it:
\be
R=-8\pi G\Biggl[\f{3\sqrt{1-q a^2}-2\sqrt{1-q
W^2}}{\sqrt{U}}\Biggr]\,.
\label{eq10}
\ee

Let us show now that in the RTG, when dealing with solutions of the
\S~ type, the situation is drastically different due to the
repulsion
force, which arises because of the stop in slowing down of the time. 

The same mechanism of the ``selflimitation of the field'', which led,
in the RTG [3-4], to inequality (\ref{eq3}) for
the external Schwar\-zsc\-hild solution, leads to inequality of the
type (\ref{eq5})
for the internal \S~ solution. RTG equations for the metric defined
by
the interval 
\be
ds^2=c^2U(W)dt^2-V(W)\acute{r}^2dW^2
-W^2(d\theta^2+\sin^2\theta\,d\phi^2)\,,
\label{eq11}
\ee
(here  $\acute{r}=dr/dW$) assume the form [5, 6]:
\be
1-\f{d}{daw}\Big[\f{W}{V\acute{r}^2}\Big]
+\f{\,1\,}{2}\Big(\f{m_g c}{\hbar}\Big)^2
\Big[W^2-r^2+\f{W^2}{2}\Big(\f{1}{U}-\f{1}{V}\Big)\Big]
=\varkappa W^2\rho\,,
\label{eq12}
\ee
\be
1-\f{1}{V\acute{r}^2}-\f{W}{UV\acute{r}^2}{\overset{\;\,\prime}{U}}
+\f{\,1\,}{2}\Big(\f{m_g c}{\hbar}\Big)^2
\Big[W^2-r^2-\f{W^2}{2}\Big(\f{1}{U}-\f{1}{V}\Big)\Bigr]
=-\varkappa W^2\f{p}{c^2}\,,
\label{eq13}
\ee
\[
\f{d}{dW}\Biggl[\sqrt{\f{U}{V}}\,W^2\Biggr]=2r\sqrt{UV}\,\acute{r}\,.
\]

Introducing a new variable 
$Z=(UW^2)/(V\acute{r}^2)$ and adding Eqs. 
(\ref{eq12}) and  (\ref{eq13})we obtain:
\be
1-\f{1}{2UW}{\overset{\;\,\prime}{Z}}
+\f{m^2}{2}(W^2-r^2)
=\f{\,1\,}{2}\varkappa W^2\Bigl(\rho -\f{p}{c^2}\Bigr)\,.
\label{eq14}
\ee
Subtracting   (\ref{eq13}) from Eq. (\ref{eq12})we find 
\be
{\overset{\;\,\prime}{Z}}-2Z\f{{\overset{\;\,\prime}{U}}}
{U}-2\f{Z}{W}-\f{m^2}{2}W^3\Bigl(1-\f{U}{V}\Bigr)
=-\varkappa W^3\Bigl(\rho +\f{p}{c^2}\Bigr)U\,,
\label{eq15}
\ee
where $m=(m_g c)/\hbar$.

In our problem the components of the energy-momentum tensor of matter
are 
\[
T_0^0=\rho,\quad T_1^1=T_2^2=T_3^3=-\f{p(W)}{c^2}\,.
\]
The equation of matter
\[
\nabla_\nu (\sqrt{-g}\,T_\mu^\nu)
=\pa_\nu (\sqrt{-g}\,T_\mu^\nu)
+\f{\,1\,}{2}\sqrt{-g}\,T_{\sigma\nu}\pa_\mu g^{\sigma\nu}=0
\]
for the given problem reduces to the following form
\be
\f{1}{c^2}\f{dp}{dW}=-\Bigl(\rho +\f{p}{c^2}\Bigr)
\f{1}{2U}\f{dU}{dW}\,.
\label{eq16}
\ee
As the pressure grows towards the center of the ball, this leads to
the inequality
\be
\f{dU}{dW}>0\,,
\label{eq17}
\ee
which means that the function $U$ decreases towards  the center of
the
ball, and hence the run of time slows down in compare with that of an
inertial system. 

Due to the  \textit{constancy} of the pressure, $\rho$, (\ref{eq16})
is readily solved:
\be
\rho+\f{p}{c^2}=\f{\alpha}{\sqrt{U}}\,.
\label{eq18}
\ee
Comparing (\ref{eq9}) and (\ref{eq18}) one finds the constant
$\alpha$ 
\be
\alpha =\rho\sqrt{1-q a^2}\,.
\label{eq19}
\ee 
If we assume that
\[
m^2(W^2-r^2)\ll 1,\quad (U/V)\ll 1\,,
\]
and introduce a new variable $y=W^2$, 
then (\ref{eq14}) and (\ref{eq15}) take the form:
\be
{\overset{\;\,\prime}{Z}}=U(1-3qy)
+\f{\alpha\varkappa}{2}y\sqrt{U}\,,
\label{eq20}
\ee
\be
\sqrt{U}{\overset{\;\,\prime}{Z}}-\f{\,1\,}{y}Z\sqrt{U}
-4Z(\sqrt{U})^\prime
+\f{\alpha\varkappa}{2}y U
-\f{m^2}{4}y\sqrt{U}=0\,.
\label{eq21}
\ee
Here and below ${\overset{\;\,\prime}{Z}}=dZ/dy$.

When analysing the external spherically symmetric \S\\ solution [3,4]
we have found that due to an effective repulsive force the metric
coefficient $U$ that defined the time run slowing down in compare
with an inertial run, did not vanish even in a strong gravitational
field. 

That is why in what follows we will investigate the behaviour of the
solution to this equations at small values of 
$y$. If the graviton mass is zero then it follows from (\ref{eq7})
for small $y$ that 
\be
\sqrt{U}\simeq\f{\,1\,}{2}(3\sqrt{1-q a^2}-1)
+\f{qy}{4}+\f{1}{16}q^2y^2\,.
\label{eq22}
\ee

From this one can see, that the function $\sqrt{U}$  for the internal
\S~ solution can vanish if 
\be
3\sqrt{1-q a^2}=1\,,
\label{eq23}
\ee
what leads to the infinite value both of the pressure $p$ and the
scalar curvature $R$ at the center of the ball.

As for the non-zero graviton rest mass Eqs. (\ref{eq20}--\ref{eq21})
stop the time-slowing-down process, one can naturally expect that
inequality  (\ref{eq23}) cannot take place in the physical (real)
domain of values of the function 
$\sqrt{U}$. We will search, on the basis of 
(\ref{eq22}), for a solution to equations 
(\ref{eq20}--\ref{eq21})   $\sqrt{U}$ in the form
\be
\sqrt{U}=\beta
+\f{qy}{4}+\f{1}{16}q^2y^2\,,
\label{eq24}
\ee
where $\beta$  is an unknown constant, which has to be  defined
making use of Eqs.(\ref{eq20}--\ref{eq21}). 
Substituting expression (24) and (25), and integrating,
we find
\be
Z=\beta^2 y+\f{y^2}{2}
\Bigl(\f{\beta q}{2}-3\beta^2 q+\f{\alpha\varkappa\beta}{2}\Bigr)
+\f{y^3}{3}\Bigl[\f{q^2}{8}\Bigl(\beta+\f{\,1\,}{2}\Bigr)
-\f{3\beta}{2}q^2+\f{\alpha\varkappa q}{8}\Bigl]\,.
\label{eq25}
\ee
Taking into expressions (\ref{eq24})  and (\ref{eq25}), and
neglecting small terms of order  $(my)^2$, we obtain the following
equation for $\beta$
\be
2\beta^2 q +\beta (q-\alpha\varkappa)+m^2/3=0\,.
\label{eq26}
\ee
It is instructive to note that the term containing $y^2$ 
has the following form:
\[
-\f{qy^2}{48}\bigl\{7[2\beta^2
q+\beta\,(q-\alpha\varkappa)]+3m^2\bigr\}\,.
\]
One can, with the use of Eq.(26), reduce it to
\[
-\f{q}{72}m^2 y^2\,.
\]
Taking  into consideration that, by definition, 
\[
\alpha\varkappa -q =\f{\varkappa\rho}{3}\bigl(3\sqrt{1-q
a^2}-1\bigr)\,,
\]
we find from Eq.(\ref{eq26}) 
\be
\beta =\f{3\sqrt{1-q a^2}-1+\Bigl[
\bigl(3\sqrt{1-q a^2}-1\bigr)^2
-(8m^2)/\varkappa\rho\Bigr]^{1/2}}{4}\,.
\label{eq27}
\ee

Thus, the metric coefficient $U$ defining the time-slowing-down
processes  \textit{is not zero}. 

If to put the graviton rest mass to zero, expression (\ref{eq27}),
as one should expect, coincides exactly with the last term of
expression (\ref{eq22}). From formula (\ref{eq27}) one can obtain the
minimum value of $\beta$
\be
\beta_{\min} =\Bigl(\f{m^2}{2\varkappa\rho}\Bigr)^{1/2}.
\label{eq28}
\ee
The quantity $\beta$ in the function $\sqrt{U}$ defines the limit for
the time-slowing-down process by the gravitational field of the ball.
This means that further slowing down of the time run by the
gravitational field is \textit{impossible}. That is why the scalar
curvature defined by expression (\ref{eq10}) will be, as distinct
from the General Theory of Relativity (GTR), finite everywhere. So,
the very gravitational field stops
the time-slowing-down process due to the non-zero graviton mass. 

According to (27) equality (\ref{eq23}) is \textit{impossible} due to
the non-zero graviton mass, because the following inequality takes
place:
\be
3\sqrt{1-q a^2}-1\geq
2\sqrt{2}\Bigl(\f{m^2}{\varkappa\rho}\Bigr)^{1/2}\,.
\label{eq29}
\ee
Taking into account the definition 
\[
q a^2=W_g/a\,,
\]
we find on the basis of inequality  (\ref{eq29})  for
$\varkappa\rho\gg m^2$
\be
a\geq\f{\,9\,}{8}W_g\Biggl(1+\sqrt{\f{m^2}{2\varkappa\rho}}\,\Biggr)\
,.
\label{eq30}
\ee
This bound for the body radius, which arises when studying the
internal solutions, is stronger than bound (\ref{eq29}), obtained in
[3,4], in the course of the analysis of the external solution.
Inequality (\ref{eq30}), as we see, follows directly from theory,
while in  the GTR inequality (\ref{eq5}) is specially introduced to
avoid
an infinite pressure inside the body. 

From (\ref{eq18}) and (\ref{eq19}) we find the pressure:
\[
\f{p}{c^2}=\f{-\rho\sqrt{U}+\rho\sqrt{1-q a^2}}{\sqrt{U}}\,.
\]
With account of equality (\ref{eq28}) we obtain the maximum pressure
at the center of the ball 
\[
\f{p}{c^2}\simeq\rho\Bigl[\f{2\varkappa\rho}{m^2}(1-q
a^2)\Bigr]^{1/2}\,.
\]
The pressure at the center of the ball is finite, while in the GTR,
due
to (\ref{eq2}), it is infinite.

The presence, in the Relativistic Theory of Gravitation, of the 
\textit{effective repulsive force}, which arises in  strong
gravitational
fields, differs it  the essence from Einstein's GTR and Newton's 
theory of
gravitation in which only \textit{attractive forces} rule. In the
field theory of gravitation the presence of the non-zero graviton
mass and the fundamental  property to stop the time run
slowing down process  lead to the
fact that the \textit{gravitational force} may be not only an
\textit{attractive force}, but at some circumstances  (in strong
fields) even an \textit{effectively repulsive} one. The effective
repulsive force stops the time run slowing down process by the
gravitational field. The gravitational field, in the main, cannot
stop the time run of a physical process because it possesses a
fundamental property of \textit{self-restriction}.

Namely this property of the gravitational field, excluding a
possibility of the black holes formation as non-physical objects,
drastically changes the picture of the matter evolution as compared
with the GTR.

\end{document}